\documentclass[a4paper,11pt]{article}

\usepackage{jheppub} 
\usepackage{graphicx}
\usepackage{amsmath,amssymb,amsthm,bm,bigdelim,multirow}
\usepackage{color}
\usepackage{hyperref}
\usepackage{dsfont}

\newcommand{\kslash}{k\kern-1ex /}
\newcommand{\pslash}{p\kern-1ex /}
\newcommand{\qslash}{q\kern-1ex /}
\newcommand{\lslash}{l\kern-1ex /}
\newcommand{\sslash}{s\kern-1ex /}
\newcommand{\Dslash}{D\kern-1.2ex /}

\newcommand{\beqa}{\begin{eqnarray}}
\newcommand{\eeqa}{\end{eqnarray}}

\newcommand{\bd}{\begin{description}}
\newcommand{\ed}{\end{description}}

\newcommand{\ben}{\begin{eqnarray}}
\newcommand{\een}{\end{eqnarray}}

\def\lsim{\raise0.3ex\hbox{$<$\kern-0.75em\raise-1.1ex\hbox{$\sim$}}}
\def\gsim{\raise0.3ex\hbox{$>$\kern-0.75em\raise-1.1ex\hbox{$\sim$}}}
\def\simgt{\rlap{\lower 3.5 pt\hbox{$\mathchar \sim$}}\raise 2.0pt \hbox {$>$}}
\def\simlt{\rlap{\lower 3.5 pt\hbox{$\mathchar \sim$}}\raise 2.0pt \hbox {$<$}}

\begin{document}

\title{Tensor renormalization group study of (3+1)-dimensional $\mathds{Z}_2$ gauge-Higgs model at finite density}

\author[a]{Shinichiro Akiyama,}
	\affiliation[a]{Graduate School of Pure and Applied Sciences, University of Tsukuba, Tsukuba, Ibaraki
    305-8571, Japan}
    	\emailAdd{akiyama@het.ph.tsukuba.ac.jp}

  	\author[b]{Yoshinobu Kuramashi}
  	\affiliation[b]{Center for Computational Sciences, University of Tsukuba, Tsukuba, Ibaraki
    305-8577, Japan}
  	\emailAdd{kuramasi@het.ph.tsukuba.ac.jp}

\abstract{
    We investigate the critical endpoints of the (3+1)-dimensional $\mathds{Z}_2$ gauge-Higgs model at finite density together with the (2+1)-dimensional one at zero density as a benchmark using the tensor renormalization group method.
  We focus on the phase transition between the Higgs phase and the confinement phase at finite chemical potential along the critical end line. 
In the (2+1)-dimensional model, the resulting endpoint is consistent with a recent numerical estimate by the Monte Carlo simulation. In the (3+1)-dimensional case, however, the location of the critical endpoint shows disagreement with the known estimates by the mean-field approximation and the Monte Carlo studies.
This is the first application of the tensor renormalization group method to a four-dimensional lattice gauge theory and a key stepping stone toward the future investigation of the phase structure of the finite density QCD.
}
\date{\today}

\preprint{UTHEP-769, UTCCS-P-143}

\maketitle

\section{Introduction}
\label{sec:intro}

The tensor renormalization group (TRG) method \footnote{In this paper, the ``TRG method" or the ``TRG approach" refers to not only the original numerical algorithm proposed by Levin and Nave \cite{Levin:2006jai} but also its extensions \cite{PhysRevB.86.045139,Shimizu:2014uva,Sakai:2017jwp,Adachi:2019paf,Kadoh:2019kqk,Akiyama:2020soe,PhysRevB.105.L060402}.} was originally proposed to study two-dimensional (2$d$) classical spin systems in the field of condensed matter physics~\cite{Levin:2006jai}. Although the TRG method was known to have several advantages over the Monte Carlo method, it was not straightforward to apply it to particle physics, where we have to treat various theories consisting of the scalar, gauge, and fermion fields on the (3+1)$d$ space-time. At the initial stage of the study of particle physics with the TRG method, we have focused on developing an efficient method to treat the scalar, gauge, and fermion fields and verifying the following advantages of the TRG method employing the lower-dimensional models: (i) no sign problem \cite{Shimizu:2014uva,Shimizu:2014fsa,Shimizu:2017onf,Takeda:2014vwa,Kadoh:2018hqq,Kadoh:2019ube,Kuramashi:2019cgs}, (ii) logarithmic computational cost on the system size, (iii) direct manipulation of the Grassmann variables \cite{Shimizu:2014uva,Sakai:2017jwp,Yoshimura:2017jpk}, (iv) evaluation of the partition function or the path-integral itself.
Recently, the authors and their collaborators have successfully applied the TRG method to analyze the phase transitions of the (3+1)$d$ complex $\phi^4$ theory at finite density~\cite{Akiyama:2020ntf}, the (3+1)$d$ real $\phi^4$ theory~\cite{Akiyama:2021zhf}, and the (3+1)$d$ Nambu$-$Jona-Lasinio (NJL) model at high density and very low temperature~\cite{Akiyama:2020soe}. 
From these previous studies, it is shown that the TRG method efficiently works to investigate the (3+1)$d$ scalar field theories with some field regularization technique and the method allows us to directly evaluate the path integral of the (3+1)$d$ lattice fermions. The next step should be to couple the gauge fields to the matters.

\begin{figure}[htbp]
	\centering
	\includegraphics[width=0.7\hsize]{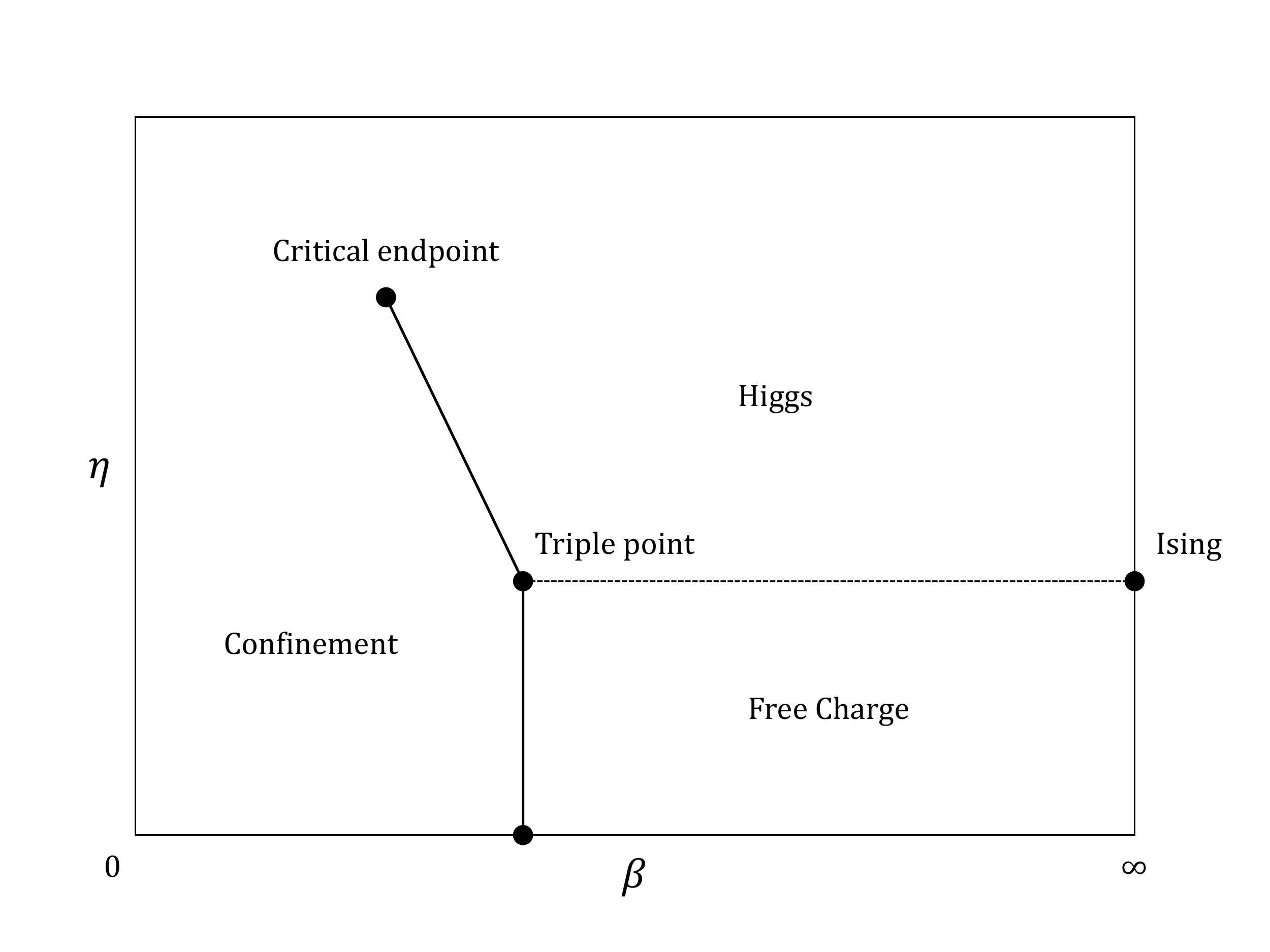}
 	\caption{Schematic phase diagram of (3+1)$d$ $\mathds{Z}_2$ gauge-Higgs model at $\mu=0$. The existence of the critical endpoint is established analytically~\cite{Balian:1974ir,Fradkin:1978dv,Brezin:1981zs} and numerically~\cite{Creutz:1979he,Jongeward:1980wx,Baig:1987ka,Blum:1998sv}. $\beta$-axis denotes the inverse gauge coupling and $\eta$ represents the spin-spin coupling. Solid lines show the first-order phase transition and dotted line is the second-order transition line~\cite{Filk:1986ds}. The pure $\mathds{Z}_{2}$ gauge theory is characterized by $\eta=0$ and the limit $\beta\to\infty$ is equivalent to the Ising model.}
  	\label{fig:phasedgm}
\end{figure}

Toward this goal, we investigate the phase structure, particularly the location of the critical endpoint\footnote{Although we mainly focus on the critical endpoint in this model, there are other interesting parameter regimes and corresponding discussions in three dimensions such as an emergent universality class at the multi-critical point ~\cite{Somoza:2020jkq,bonati2021multicritical} or possible higher-order transitions~\cite{Grady:2021fax}.}, of the (3+1)$d$ $\mathds{Z}_2$ gauge-Higgs model at finite density in this paper. 
So far, the TRG analyses on the gauge theories have been limited to the (1+1)$d$ systems~\cite{Shimizu:2014uva,Shimizu:2014fsa,Shimizu:2017onf,Unmuth-Yockey:2018ugm,Kuramashi:2019cgs,Bazavov:2019qih,Fukuma:2021cni,Hirasawa:2021qvh} and (2+1)$d$ ones~\cite{Dittrich:2014mxa,Kuramashi:2018mmi,Unmuth-Yockey:2018xak}. This study is the first application of the TRG method to a (3+1)$d$ lattice gauge theory. 
The existence of the critical endpoint in the phase diagram of $\mathds{Z}_2$ gauge-Higgs model is established both by the analytical discussions~\cite{Balian:1974ir,Fradkin:1978dv,Brezin:1981zs} and by the Monte Carlo simulations~\cite{Creutz:1979he,Jongeward:1980wx,Baig:1987ka,Blum:1998sv}, though the precise location of the critical endpoint does not seem to have been identified.
Figure~\ref{fig:phasedgm} shows a sketch of the phase diagram for the (3+1)$d$ $\mathds{Z}_2$ gauge-Higgs model at vanishing density~\cite{Balian:1974ir,Fradkin:1978dv,Creutz:1979he,Jongeward:1980wx,Brezin:1981zs,Filk:1986ds,Baig:1987ka,Blum:1998sv}. We focus on the phase transition between the Higgs phase and the confinement phase at a finite chemical potential $\mu$ along the critical end line. 
Although $\mathds{Z}_2$ gauge-Higgs model does not suffer from the sign problem even at finite density, this work is motivated by the preparation for the future investigation of the critical endpoint in the finite density QCD.  
Before studying the model in (3+1) dimensions, we firstly make a benchmark test employing the (2+1)$d$ $\mathds{Z}_2$ gauge-Higgs model whose phase structure shares the similar features with the (3+1)$d$ case. 
In the (2+1)$d$ case, the location of the critical endpoint at $\mu=0$ is consistently reproduced with recent Monte Carlo studies~\cite{PhysRevB.82.085114,Somoza:2020jkq,bonati2021multicritical}. On the other hand, in the (3+1)$d$ case, we observe a discrepancy between the TRG result and those obtained by the mean-field approximation~\cite{Brezin:1981zs} and the Monte Carlo studies~\cite{Creutz:1979he}. 

This paper is organized as follows. In Sec.~\ref{sec:method}, we define the $\mathds{Z}_2$ gauge-Higgs model at finite density on a lattice in arbitrary dimension and explain how to construct its tensor network representation. We present the results of the benchmark test using the (2+1)$d$ $\mathds{Z}_2$ gauge-Higgs model at $\mu=0$ in Sec.~\ref{sec:results}. 
After that, we determine the critical endpoints at $\mu=0$, $1$, $2$ in the (3+1)$d$ model and discuss how they are shifted by the effect of finite $\mu$. 
Section~\ref{sec:summary} is devoted to summary and outlook.

\section{Formulation and numerical algorithm}
\label{sec:method}

\subsection{($d$+1)-dimensional $\mathds{Z}_2$ gauge-Higgs model at finite density}
\label{subsec:action}

We consider the partition function of the $\mathds{Z}_2$ gauge-Higgs model at finite density on an isotropic hypercubic lattice $\Lambda_{d+1}=\{(n_1,\dots,n_{d+1})\ \vert n_{\nu}=1,\dots ,L\}$ whose volume is equal to $V=L^{d+1}$. The lattice spacing $a$ is set to $a=1$ without loss of generality. The gauge fields $U_{\nu}(n)$ ($\nu=1,\dots,d+1$) reside on the links and the matter fields $\sigma(n)$ are on the sites. Both variables $U_{\nu}(n)$ and $\sigma(n)$ take their values on $\mathds{Z}_{2}=\{\pm1\}$.
The action $S$ is defined as
\begin{align}
\label{eq:action}
	S=&-{\beta}\sum_{n\in\Lambda_{d+1}}\sum_{\nu>\rho} U_{\nu}(n)U_{\rho}(n+\hat{\nu})U_{\nu}(n+\hat{\rho})U_{\rho}(n)\nonumber\\
	&-\eta \sum_{n}\sum_{\nu}\left[{\rm e}^{\mu\delta_{\nu,d+1}}\sigma(n) U_{\nu}(n) \sigma(n+{\hat\nu})+{\rm e}^{-\mu\delta_{\nu,d+1}}\sigma(n) U_{\nu}(n-{\hat \nu})\sigma(n-{\hat\nu})\right],
\end{align}
where 
$\beta$ is the inverse gauge coupling, $\eta$ is the gauge-invariant spin-spin coupling and $\mu$ is the chemical potential. This parametrization follows Ref.~\cite{Gattringer:2012jt}. We employ the periodic boundary conditions for both the gauge and matter fields in all the directions.
The partition function is then given by 
\begin{align}
\label{eq:Z}
	Z=\left(\prod_{n,\nu}\sum_{U_{\nu}(n)=\pm1}\right)\left(\prod_{n}\sum_{\sigma(n)=\pm1}\right){\rm e}^{-S},
\end{align}
where the sum is taken over all possible field configurations. Since $\sigma(n)\in\mathds{Z}_{2}$, one is allowed to choose the so-called unitary gauge \cite{Creutz:1979he}, which eliminates the matter field $\sigma(n)$ by redefining the link variable $U_{\nu}(n)$ via
\begin{align}
	\sigma(n) U_{\nu}(n) \sigma(n+{\hat\nu})\mapsto U_{\nu}(n).
\end{align}
With the unitary gauge, Eq.~\eqref{eq:action} is reduced to be
\begin{align}
	S&=-{\beta}\sum_{n\in\Lambda_{d+1}}\sum_{\nu>\rho} U_{\nu}(n)U_{\rho}(n+\hat{\nu})U_{\nu}(n+\hat{\rho})U_{\rho}(n)-2\eta\sum_{n}\sum_{\nu}\cosh\left(\mu\delta_{\nu,d+1}\right)U_{\nu}(n),
\end{align}
whose partition function is 
\begin{align}	
\label{eq:Z_ug}
	Z=\left(\prod_{n,\nu}\sum_{U_{\nu}(n)=\pm1}\right){\rm e}^{-S},
\end{align}
instead of Eq.~\eqref{eq:Z}.

\subsection{Tensor network representation of lattice gauge fields}
\label{subsec:tn-rep}

\begin{figure}[htbp]
	\centering
	\includegraphics[width=0.9\hsize]{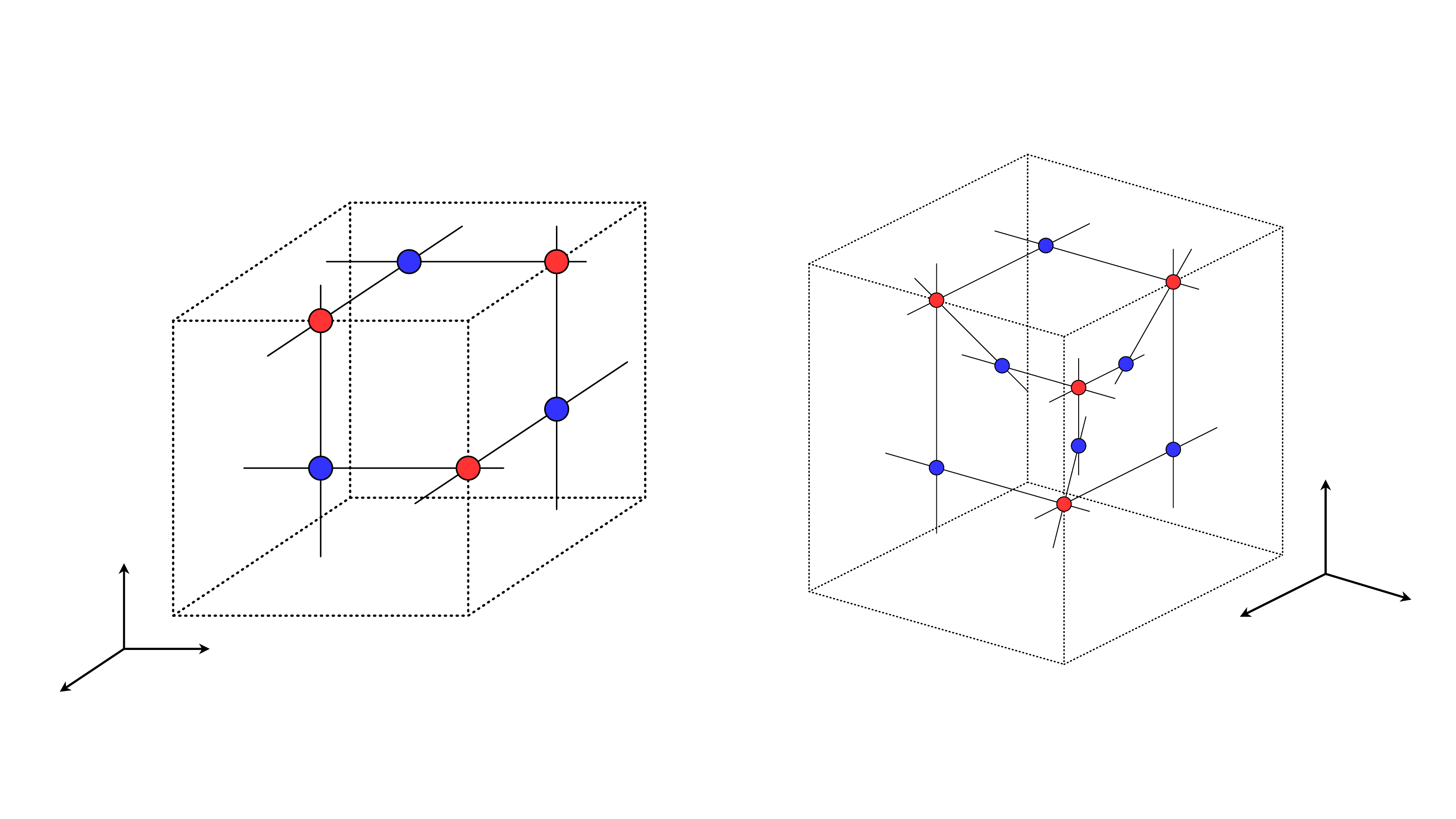}
	\caption{Structures of $\mathcal{T}_{c}$ in Eq.~\eqref{eq:tn_lgt} in $\Lambda_{2+1}$ (left) and $\Lambda_{3+1}$ (right). The lattice geometry in real space is represented by dotted lines in both cases. Red symbols show link tensors in corresponding dimensions and blue ones are plaquette tensors. Skewed directions in the four-dimensional description correspond to the forth direction in $\Lambda_{3+1}$.}
  	\label{fig:tn_lgt}
\end{figure}

Although a tensor network representation of the $\mathds{Z}_{2}$ gauge theory is constructed in Ref.~\cite{Liu:2013nsa}, which is successfully applied in the numerical calculation in Ref.~\cite{Kuramashi:2018mmi}, we introduce a little bit different way to derive a tensor network representation for $\mathds{Z}_{2}$ gauge fields on $\Lambda_{d+1}$. Firstly, we regard a local Boltzmann weight corresponding to a plaquette interaction as a four-rank tensor,
\begin{align}
	W_{U_{\nu}(n)U_{\rho}(n+\hat{\nu})U_{\nu}(n+\hat{\rho})U_{\rho}(n)}=\exp\left[\beta U_{\nu}(n)U_{\rho}(n+\hat{\nu})U_{\nu}(n+\hat{\rho})U_{\rho}(n)\right],
\end{align}
whose higher-order singular value decomposition (HOSVD) gives us
\begin{align}
\label{eq:hosvd}
	W_{U_{\nu}(n)U_{\rho}(n+\hat{\nu})U_{\nu}(n+\hat{\rho})U_{\rho}(n)}=\sum_{a,b,c,d}V_{U_{\nu}(n)a}V_{U_{\rho}(n+\hat{\nu})b}V_{U_{\nu}(n+\hat{\rho})c}V_{U_{\rho}(n),d}B_{abcd},
\end{align}
where $V$'s are unitary matrices and $B$ is the so-called core tensor. Thanks to this decomposition, we can integrate out all link variables $U_{\nu}(n)$'s in Eq.~\eqref{eq:Z_ug} at each link independently. As a result of this integration, we have a $2d$-rank tensor $A$ at each link according to
\begin{align}
\label{eq:pure_link}
	A_{m_{1}\cdots m_{d}m'_{1}\cdots m'_{d}}=\sum_{U_{\nu}(n)}\left(\prod_{j=1}^{d}V_{U_{\nu}(n)m_{j}}V_{U_{\nu}(n)m'_{j}}\right)M_{U_{\nu}(n)},
\end{align}
with
\begin{align}
	M_{U_{\nu}(n)}=\exp\left[2\eta\cosh\left(\mu\delta_{\nu,d+1}\right)U_{\nu}(n)\right].
\end{align}
Since the tensor $A$ lives on the each link $\ell=(n,\nu)$, we call it the link tensor, denoting $A_{\ell;m_{1}\cdots m_{d}m'_{1}\cdots m'_{d}}$. Similarly, the core tensor $B$ is located on each plaquette  $\square=(n,\nu,\rho)$, so we call it the plaquette tensor, denoting $B_{\square;abcd}$. Therefore, we have a tensor network representation such as
\begin{align}
	Z={\rm tTr}\left[\left(\prod_{\ell}A_{\ell}\right)\left(\prod_{\square}B_{\square}\right)\right].
\end{align}
However, conventional TRG algorithms usually consider a tensor network representation described just by a single kind of tensor located at each lattice site. We now follow the asymmetric formulation provided in Ref.~\cite{Liu:2013nsa}, which allows us to have a uniform tensor network representation of Eq.~\eqref{eq:Z_ug} as in the form of
\begin{align}
\label{eq:tn_lgt}
	Z={\rm tTr}\left[\prod_{c\in\Lambda_{d+1}}\mathcal{T}_{c}\right],
\end{align}
where a basic cell in the lattice $\Lambda_{d+1}$ is represented by $c$ and $\mathcal{T}_{c}$ is a $2(d+1)$-rank tensor generated by $(d+1)$ pieces of link tensors and $d(d+1)/2$ pieces of plaquette tensors. Figure~\ref{fig:tn_lgt} illustrates structures of $\mathcal{T}_{c}$ in higher dimensions. 

It is worth noting that the above derivation of tensor network representation is easily applicable to $\mathds{Z}_{N}$ gauge theories with general $N$. In addition, one can introduce a lower-rank approximation via the HOSVD of plaquette weights as in Eq.~\eqref{eq:hosvd}. Although our target in this paper corresponds to the case with $N=2$ and no HOSVD-based approximation is necessary to derive the tensor network representation, this treatment is practically useful to compress the size of $\mathcal{T}_{c}$ when we consider $N\gg2$.

\subsection{A remark on the TRG algorithm}

In this work, we employ the anisotropic TRG (ATRG) algorithm~\cite{Adachi:2019paf} to evaluate Eq.~\eqref{eq:tn_lgt}. Both in (2+1)- and (3+1)-dimensional cases, the ATRG is parallelized according to Refs.~\cite{Akiyama:2020Dm,Akiyama:2020ntf}. As a singular value decomposition (SVD) algorithm in the bond-swapping procedure explained in Refs.~\cite{Adachi:2019paf,Oba:2019csk}, the randomized SVD (RSVD) is applied choosing $p=2D$ and $q=D$, where $p$ is the oversampling parameter, $q$ is the iteration numbers of QR decompositions in the RSVD, and $D$ is the bond dimension in the ATRG algorithm.

\section{Numerical results} 
\label{sec:results}

\subsection{Study of the (2+1)-dimensional model as a benchmark}
\label{subsec:(2+1)d}

The partition function of Eq.~\eqref{eq:Z_ug} is evaluated using the parallelized ATRG algorithm on lattices with the volume $V=L^3$ with the periodic boundary condition in all the directions. 
In the following, all the results are calculated setting $D=48$ on a lattice whose volume is $1024^{3}$. Up to $V=1024^{3}$, the TRG computation converges with respect to the system size and allows us to access the thermodynamic limit.

\begin{figure}[htbp]
	\centering
	\includegraphics[width=0.8\hsize]{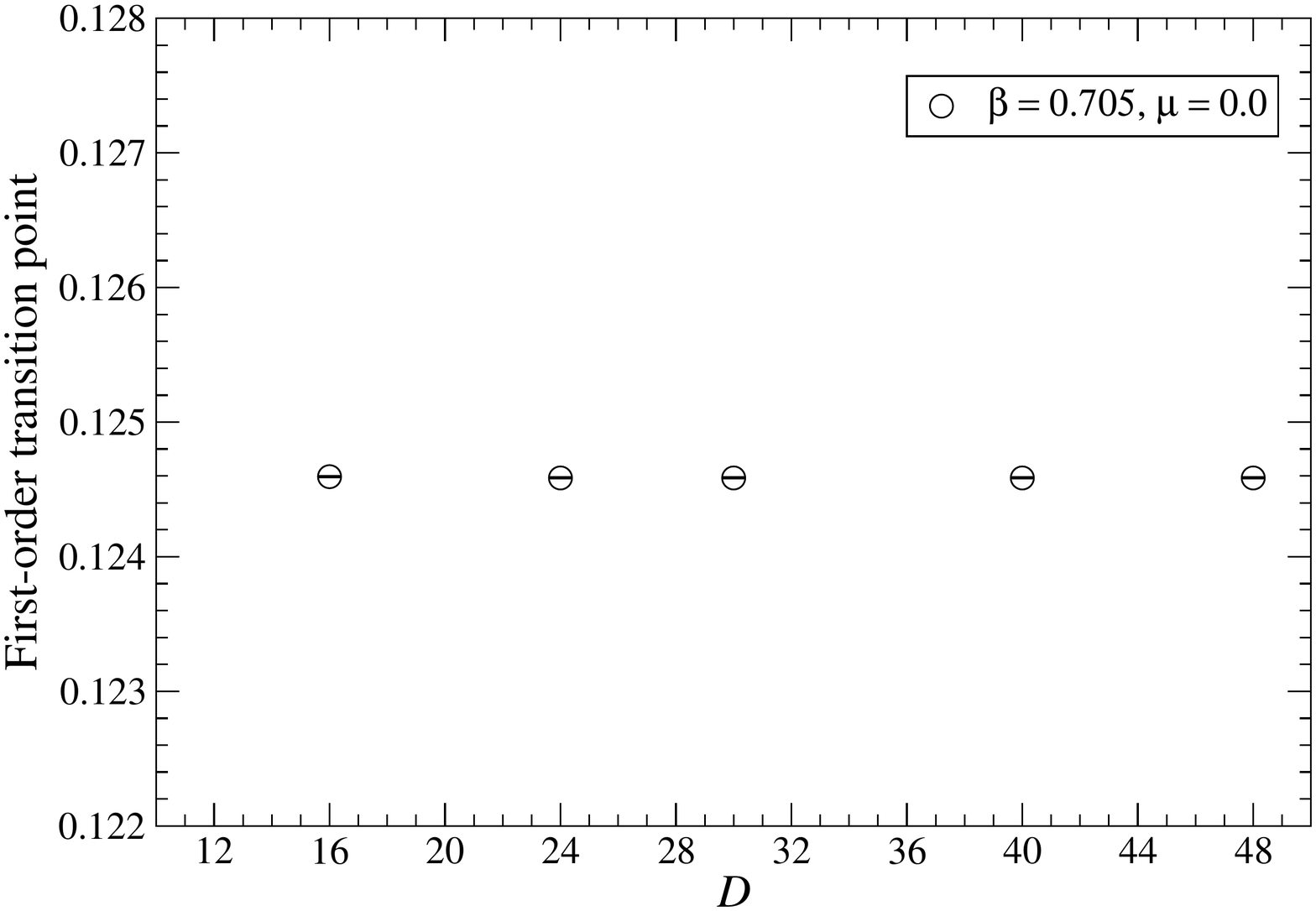}
	\caption{Convergence behavior of the transition point of $\eta$ as a function of $D$ at $\beta=0.705$ with vanishing $\mu$. Error bars are all within the symbols.}
  	\label{fig:3d_eta_D}
\end{figure}

\begin{figure}[htbp]
	\centering
	\includegraphics[width=0.8\hsize]{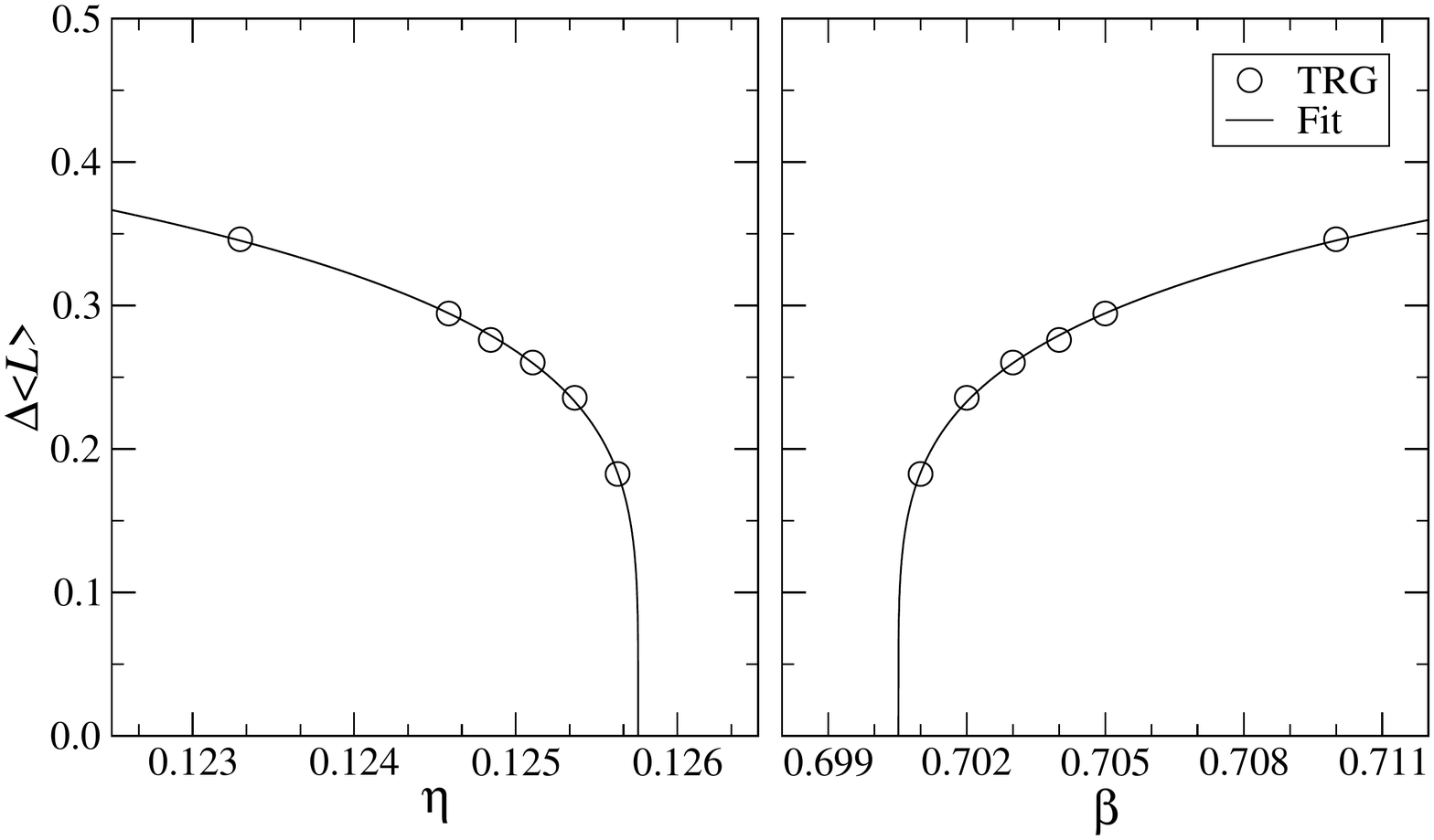}
	\caption{(Left) Fit of $\Delta \langle L\rangle$ at $\mu=0$ as a function of $\eta$. (Right) Fit of $\Delta \langle L\rangle$ at $\mu=0$ as a function of $\beta$.}
  	\label{fig:delta_3d_mu0_fit}
\end{figure}

\begin{figure}[htbp]
	\centering
	\includegraphics[width=0.8\hsize]{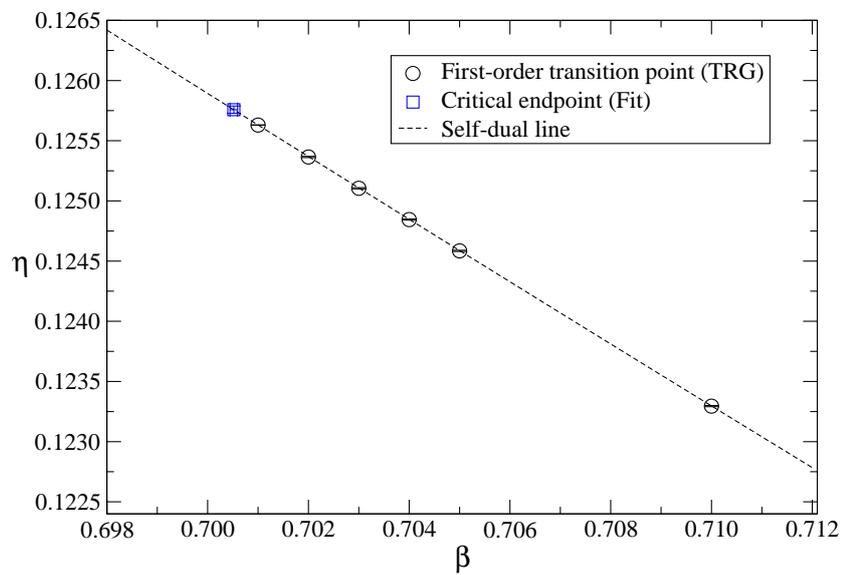}
	\caption{Phase diagram in the vicinity of the critical endpoint. Circles show the first-order phase transition points and the critical endpoint is denoted by the blue symbol. Error bars are all within the symbols. Dotted line is the self-dual line defined by Eq.~\eqref{eq:s_dual} at which all transition points are located.}
  	\label{fig:dl}
\end{figure}

We determine the critical endpoint  $(\beta_{\rm c},\eta_{\rm c})$ at $\mu=0$, where the first-order phase transition line terminates. We employ the average link defined by
\begin{align}
\label{eq:def_al}
	\langle L\rangle=\frac{1}{(d+1)V}\frac{\partial \ln Z}{\partial(2\eta)}
\end{align}
to detect the first-order phase transition. The factor $(d+1)V$ corresponds to the number of links in $\Lambda_{d+1}$ with the periodic boundary condition. We evaluate $\langle L\rangle$ with the impurity tensor method. The tensor network representation of $\langle L\rangle$ is presented in appendix~\ref{sec:impure}.
Figure~\ref{fig:link_3d_mu=0} shows the $\eta$ dependence of the average link with the several choices of $\beta\in[0.700,0.710]$. We observe clear gaps of $\langle L\rangle$ with $\beta\in[0.701,0.710]$, though it is difficult to identify such a gap at $\beta=0.700$. The value of the gap of $\langle L\rangle$, which is denoted by $\Delta \langle L\rangle$, is listed in Table~\ref{tab:link_3d}, together with the corresponding first-order transition point $(\beta,\eta)$. Although $\Delta \langle L\rangle$ is evaluated just by $\langle L\rangle(\eta=\eta_{+})-\langle L\rangle(\eta=\eta_{-})$, where $\eta_{+}$ and $\eta_{-}$ are chosen from different phases, we set $\eta_{+}-\eta_{-}={\rm O}(10^{-5})$ for $\beta\in[0.702,0.710]$ and $\eta_{+}-\eta_{-}={\rm O}(10^{-6})$ at $\beta=0.701$. The error for $\eta$ in Table~\ref{tab:link_3d} is provided by the magnitude of $\eta_{+}-\eta_{-}$. 
We have also checked the finite-$D$ effect via monitoring the $D$ dependence of the transition point $\eta$ as shown in Fig.~\ref{fig:3d_eta_D}. It has converged up to the fifth decimal to $D$ and we neglect the finite-$D$ effect in the current computation.
To determine the critical endpoint $(\beta_{\rm c},\eta_{\rm c})$, we separately fit the data of $\Delta \langle L\rangle$ assuming the functions $\Delta \langle L\rangle=A(\beta-\beta_{\rm c})^p$ and $\Delta \langle L\rangle=B(\eta_{\rm c}-\eta)^q$, respectively, where $A$, $B$, $\beta_{\rm c}$, $\eta_{\rm c}$, $p$, and $q$ are the fit parameters. The fit results are depicted in Fig.~\ref{fig:delta_3d_mu0_fit} and their numerical values are presented in Table~\ref{tab:link_3d_fit}. Figure~\ref{fig:dl} summarizes the first-order transition points in Table~\ref{tab:link_3d} and the critical endpoint in Table~\ref{tab:link_3d_fit} on the $\beta$-$\eta$ plane, comparing them with the self-dual line defined by
\begin{align}
\label{eq:s_dual}
	\eta=-\frac{1}{4}\ln\tanh\beta.
\end{align}
The self duality is a special feature in the three-dimensional $\mathds{Z}_{2}$ gauge-Higgs model as demonstrated in Ref.~\cite{Balian:1974ir}. Figure~\ref{fig:dl} tells us that the first-order transitions and the critical endpoint are actually on the self-dual line as expected. Moreover, the location of the critical endpoint $(\beta_{\rm c},\eta_{\rm c})=(0.70051(7), 0.12575(3))$ is consistent with the previous result $\beta_{\rm c}\approx0.701$~\cite{Somoza:2020jkq}. 
These do assure the validity of the current TRG-based determination of the critical endpoint, which is characterized as a point with vanishing $\Delta\langle L\rangle$.

\begin{figure}[htbp]
	\centering
	\includegraphics[width=0.8\hsize]{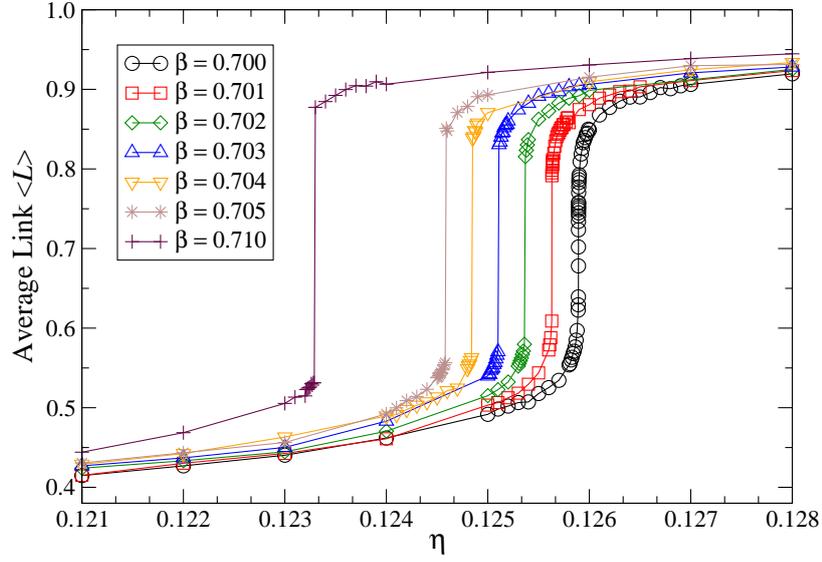}
	\caption{$\eta$ dependence of $\langle L\rangle$ at $\mu=0$ for $\beta\in[0.700,0.710]$ on a lattice whose volume is $V=1024^3$.}
  	\label{fig:link_3d_mu=0}
\end{figure}

\begin{table}[htb]
	\caption{$\Delta \langle L\rangle$ and the first-order transition points $(\beta,\eta)$ at $\mu=0$ on a $1024^3$ lattice.}
	\label{tab:link_3d}
	\begin{center}
	  	\begin{tabular}{|c|c|c|}\hline
          	\multicolumn{3}{|c|}{$\mu=0$}  \\ \hline
  		$\beta$ & $\eta$ & $\Delta \langle L\rangle$  \\ \hline
  		0.701 & 0.1256305(5) & 0.18258788 \\
		0.702 & 0.125365(5) & 0.23570012 \\
		0.703 & 0.125105(5) & 0.26027553 \\
		0.704 & 0.124845(5) & 0.27604029 \\
		0.705 & 0.124585(5) & 0.29445449 \\
  		0.710 & 0.123295(5) & 0.34614902 \\ \hline
	\end{tabular}
	\end{center}
\end{table}

\begin{table}[htb]
	\caption{Fit results for $\Delta \langle L\rangle$ at $\mu=0$ in the (2+1)$d$ case.}
	\label{tab:link_3d_fit}
	\begin{center}
	  	\begin{tabular}{|ccc|ccc|}\hline
          	\multicolumn{6}{|c|}{$\mu=0$}  \\ \hline
			$A$ & $\beta_{\rm c}$ & $p$ & $B$ & $\eta_{\rm c}$ & $q$ \\ \hline
			0.92(5) & 0.70051(7) & 0.21(1) & 1.24(8) & 0.12575(3) & 0.21(1) \\ \hline
		\end{tabular}
	\end{center}
\end{table}

\clearpage

\subsection{(3+1)-dimensional model at finite density}
\label{subsec:(3+1)d}

\begin{figure}[htbp]
	\centering
	\includegraphics[width=0.8\hsize]{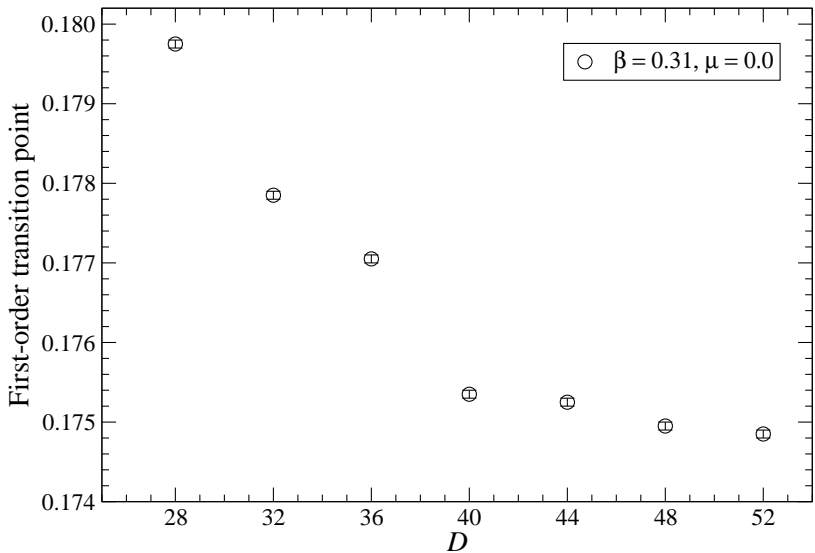}
	\caption{Convergence behavior of the transition point of $\eta$ as a function of $D$ at $\beta=0.31$ with vanishing $\mu$. Error bars are all within the symbols.}
  	\label{fig:eta_D}
\end{figure}

Now, we move on to the investigation of the model in the (3+1) dimension.
We firstly check the convergence behavior of the transition point for the bond dimension $D$. Figure~\ref{fig:eta_D} shows the transition point obtained by calculating the average link in Eq.~\eqref{eq:def_al} at $\beta=0.31$ with vanishing $\mu$. As we see below, this transition point is close to the critical endpoint. With $D>40$, we see that the finite-$D$ effect is well suppressed: the relative error between the first-order transition points with $D=48$ and $D=52$ is $0.057\%$. 
Hereafter, we present the results, fixing $D=48$, on a lattice whose volume is $32^{4}$. Although the number of lattice sites is much smaller than the (2+1)$d$ case, $V=32^{4}$ is sufficiently large to be regarded as the thermodynamic limit in the (3+1)$d$ case. The TRG calculation has converged within 20 times of iteration.

\begin{figure}[htbp]
	\centering
	\includegraphics[width=0.8\hsize]{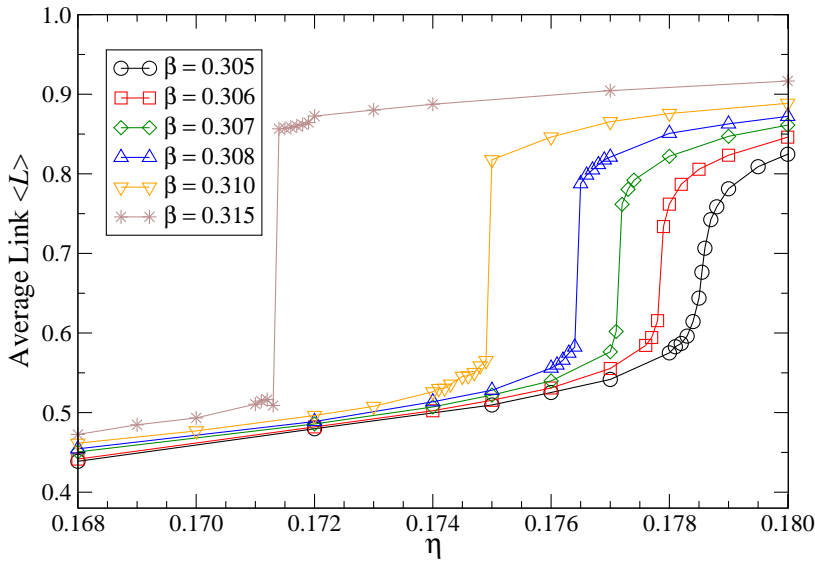}
	\caption{$\eta$ dependence of $\langle L\rangle$ at $\mu=0$ for $\beta\in[0.305,0.315]$.}
  	\label{fig:link_4d_mu=0}
\end{figure}

\begin{figure}[htbp]
  	\centering
	\includegraphics[width=0.8\hsize]{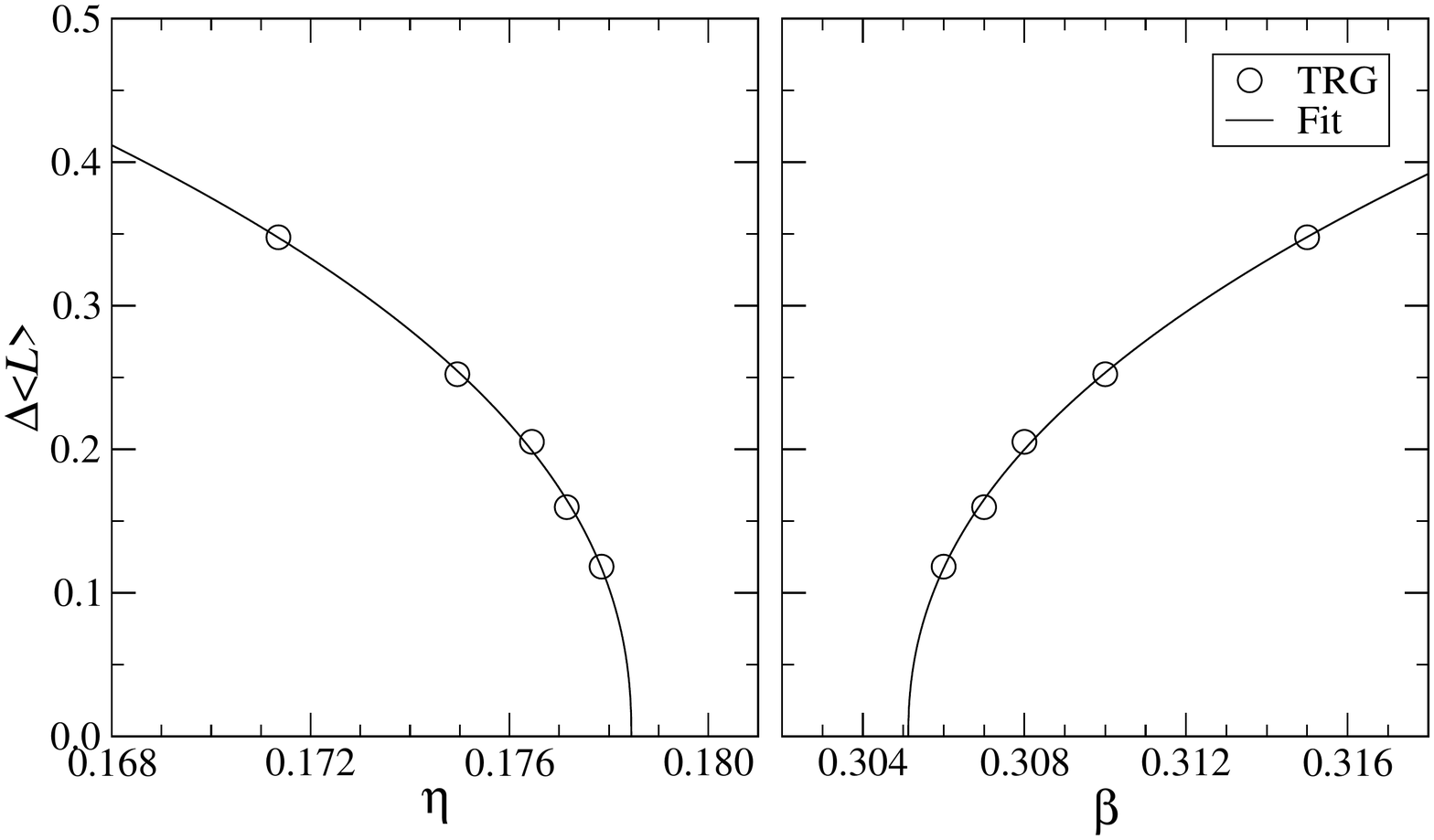}
	\caption{(Left) Fit of $\Delta \langle L\rangle$ at $\mu=0$ as a function of $\eta$. (Right) Fit of $\Delta \langle L\rangle$ at $\mu=0$ as a function of $\beta$.}
  	\label{fig:delta_4d_mu0_fit}
\end{figure}

We show the $\eta$ dependence of $\langle L\rangle$ at $\mu=0$ with the several choices of $\beta$ in Fig.~\ref{fig:link_4d_mu=0}, where the gap of $\langle L\rangle$ is clearly observed at a certain value of $\eta$ for $\beta\in[0.306,0.315]$.
As in the (2+1)$d$ case, we determine $\beta_{\rm c}$ and $\eta_{\rm c}$ separately, by fitting the data of $\Delta \langle L\rangle$ in Table~\ref{tab:link_4d}, assuming the functions $\Delta \langle L\rangle=A(\beta-\beta_{\rm c})^p$ and $\Delta \langle L\rangle =B(\eta_{\rm c}-\eta)^q$, respectively. The fit results are depicted in Fig.~\ref{fig:delta_4d_mu0_fit} and their numerical values are presented in Table~\ref{tab:link_4d_fit}. Note that $\Delta \langle L\rangle$ is estimated as in the same way to the previous (2+1)$d$ analysis, setting $\eta_{+}-\eta_{-}={\rm O}(10^{-4})$.
We obtain $(\beta_{\rm c},\eta_{\rm c})=(0.3051(2),0.1784(2))$ for the critical endpoint at $\mu=0$. 
According to the mean-field theory, the critical endpoint is located at $(\beta_{\rm c},2\eta_{\rm c})=(2/(3d),\ln(1+\sqrt{2})-\sqrt{2}/3)$, which is roughly equal to $(0.22,0.41)$~\cite{Brezin:1981zs}. The Monte Carlo simulation for this model on an $8^{4}$ lattice estimates $(\beta_{\rm c},2\eta_{\rm c})=(0.22(3),0.48(3))$~\cite{Creutz:1979he}. Our result is not consistent with these previous results. It is, however, difficult to discuss the origin of the discrepancy, because Ref.~\cite{Creutz:1979he} does not explain how to estimate the location of the critical endpoint. Additionally, it may be worth emphasizing that the current TRG computation allows us to capture a clear gap of $\langle L\rangle$ in the vicinity of transition points characterized by $\eta_{+}-\eta_{-}={\rm O}(10^{-4})$, and $\langle L\rangle$ does become smooth at $\beta=0.305$ as shown in Figure~\ref{fig:link_4d_mu=0}.

\begin{table}[htb]
	\caption{$\Delta \langle L\rangle$ and the first-order transition points $(\beta,\eta)$ at $\mu=0$, $1$, $2$. All the results are obtained with $D=48$ in the TRG method.}
	\label{tab:link_4d}
	\begin{center}
	  	\begin{tabular}{|c|c|c|}\hline
          	\multicolumn{3}{|c|}{$\mu=0$}  \\ \hline
			$\beta$ & $\eta$ & $\Delta \langle L\rangle$  \\ \hline
			0.306 & 0.17785(5) & 0.11825357  \\
			0.307 & 0.17715(5) & 0.15964584  \\
			0.308 & 0.17645(5) & 0.20518511  \\
			0.310 & 0.17495(5) & 0.25228994  \\
			0.315 & 0.17135(5) & 0.34764255  \\ \hline
         		\multicolumn{3}{|c|}{$\mu=1$}  \\ \hline
			$\beta$ & $\eta$ & $\Delta \langle L\rangle$  \\ \hline
			0.306 & 0.15895(5) & 0.16477722  \\
			0.307 & 0.15815(5) & 0.21320870  \\
			0.308 & 0.15755(5) & 0.24463033  \\
			0.309 & 0.15685(5) & 0.25522649  \\
			0.310 & 0.15625(5) & 0.28582312  \\
			0.311 & 0.15555(5) & 0.29552291  \\ \hline
           		 \multicolumn{3}{|c|}{$\mu=2$}  \\ \hline
			$\beta$ & $\eta$ & $\Delta \langle L\rangle$  \\ \hline
			0.298 & 0.12595(5) & 0.16413038  \\
			0.299 & 0.12545(5) & 0.20458404  \\
			0.300 & 0.12495(5) & 0.23911321  \\
			0.301 & 0.12445(5) & 0.26057957  \\
			0.302 & 0.12395(5) & 0.27639030  \\
			0.303 & 0.12345(5) & 0.29988375  \\ \hline
	\end{tabular}
	\end{center}
\end{table}

\begin{table}[htb]
	\caption{Fit results for $\Delta \langle L\rangle$. All the results are obtained with $D=48$ in the TRG method.}
	\label{tab:link_4d_fit}
	\begin{center}
	  \begin{tabular}{|ccc|ccc|}\hline
          \multicolumn{6}{|c|}{$\mu=0$}  \\ \hline
$A$ & $\beta_{\rm c}$ & $p$ & $B$ & $\eta_{\rm c}$ & $q$ \\ \hline
2.7(4) &  0.3051(2) & 0.44(3) & 3.0(6) & 0.1784(2) & 0.43(4) \\ \hline
          \multicolumn{6}{|c|}{$\mu=1$}  \\ \hline
$A$ & $\beta_{\rm c}$ & $p$ & $B$ & $\eta_{\rm c}$ & $q$ \\ \hline
1.1(2) &  0.3053(2) & 0.26(4) & 1.6(6) & 0.1595(3) & 0.30(7) \\ \hline
          \multicolumn{6}{|c|}{$\mu=2$}  \\ \hline
$A$ & $\beta_{\rm c}$ & $p$ & $B$ & $\eta_{\rm c}$ & $q$ \\ \hline
1.6(2) &  0.2969(2) & 0.33(3) & 2.0(4) & 0.1264(1) & 0.33(4) \\ \hline
	\end{tabular}
	\end{center}
\end{table}

\begin{figure}[htbp]
	\centering
	\includegraphics[width=0.8\hsize]{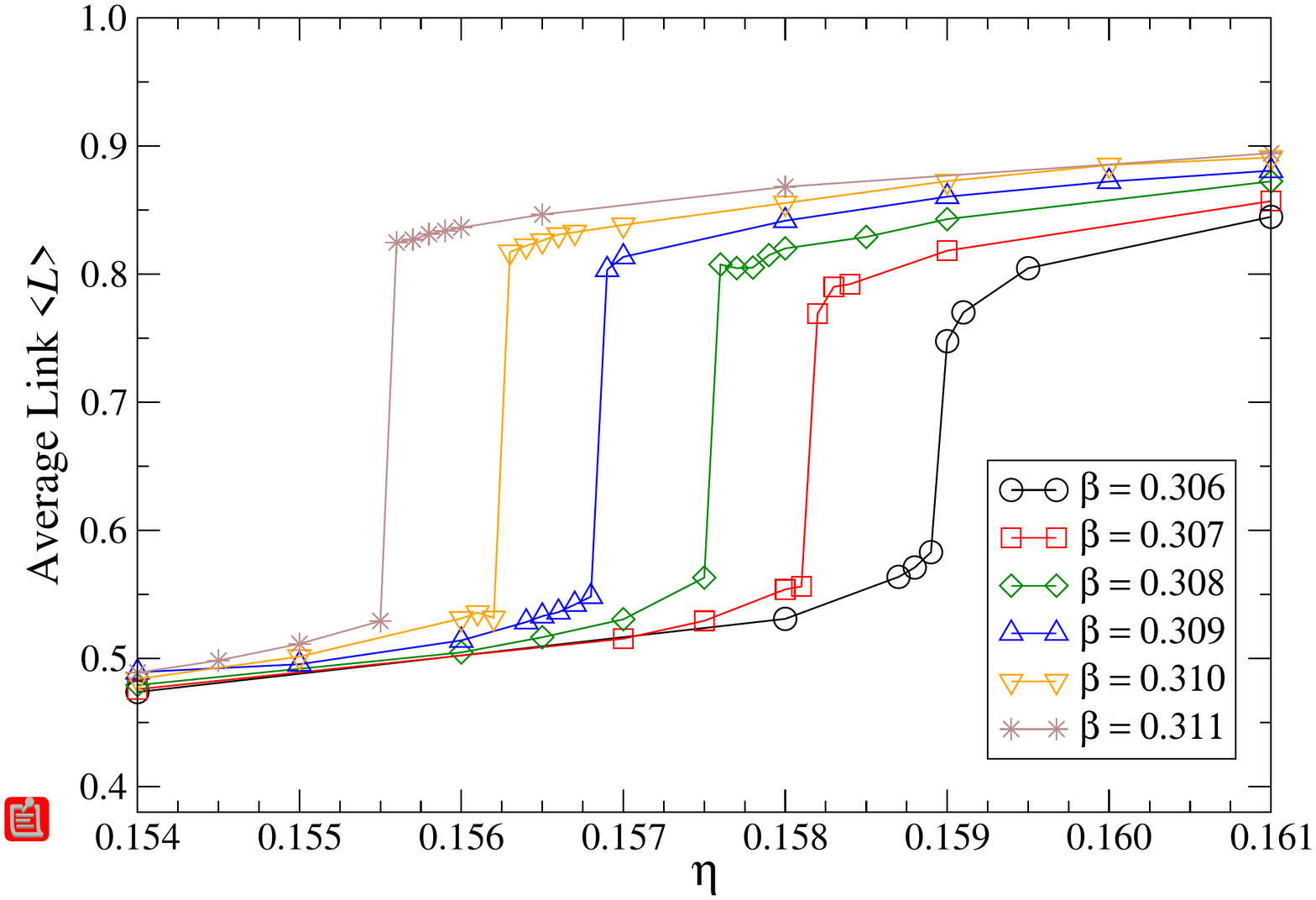}
	\caption{$\eta$ dependence of $\langle L\rangle$ at $\mu=1$ for $\beta\in [0.306,0.311]$.}
  	\label{fig:link_mu=1}
\end{figure}

\begin{figure}[htbp]
	\centering
	\includegraphics[width=0.8\hsize]{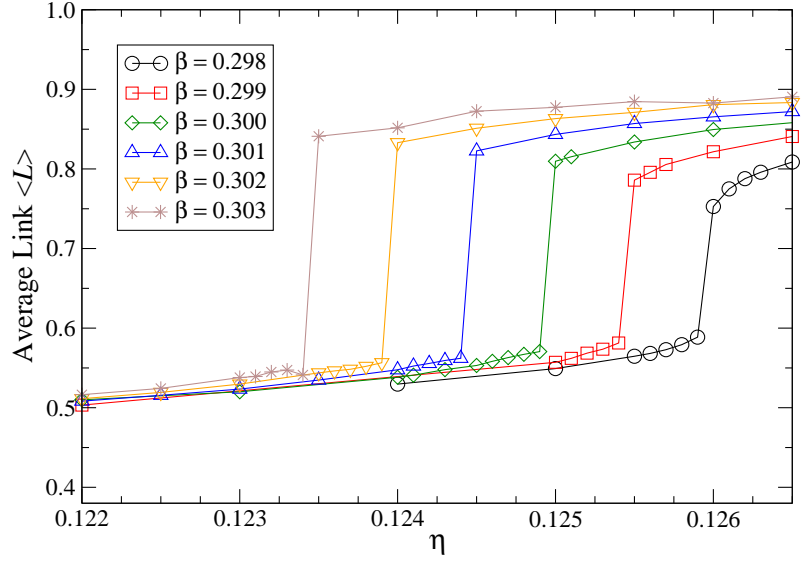}
	\caption{$\eta$ dependence of $\langle L\rangle$ at $\mu=2$ for $\beta\in [0.298,0.303]$.}
  	\label{fig:link_mu=2}
\end{figure}

\begin{figure}[htbp]
  	\centering
	\includegraphics[width=0.8\hsize]{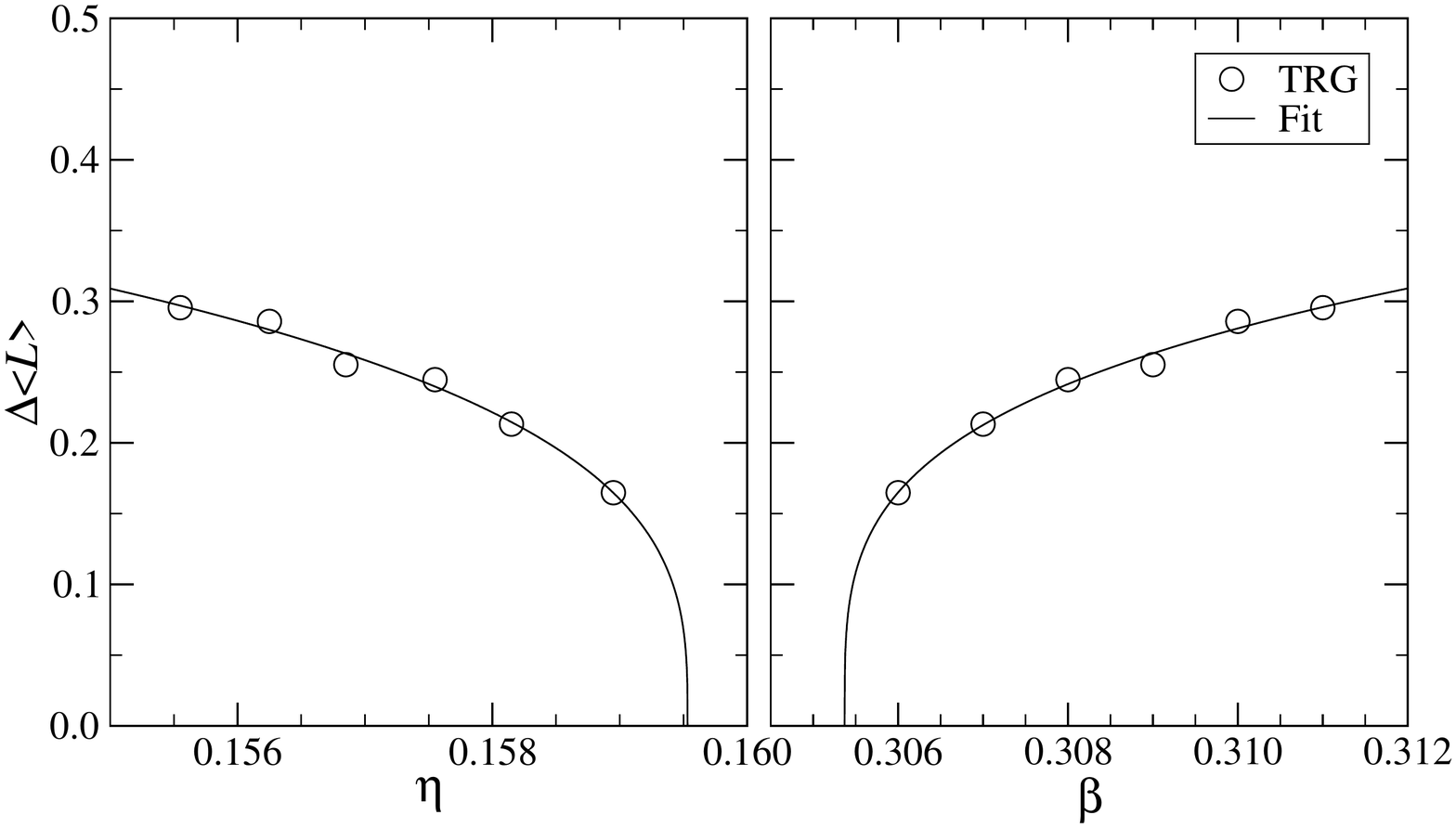}
	\caption{(Left) Fit of $\Delta \langle L\rangle$ at $\mu=1$ as a function of $\eta$. (Right) Fit of $\Delta \langle L\rangle$ at $\mu=1$ as a function of $\beta$.}
	\label{fig:delta_4d_mu1_fit}
\end{figure}

\begin{figure}[htbp]
  	\centering
	\includegraphics[width=0.8\hsize]{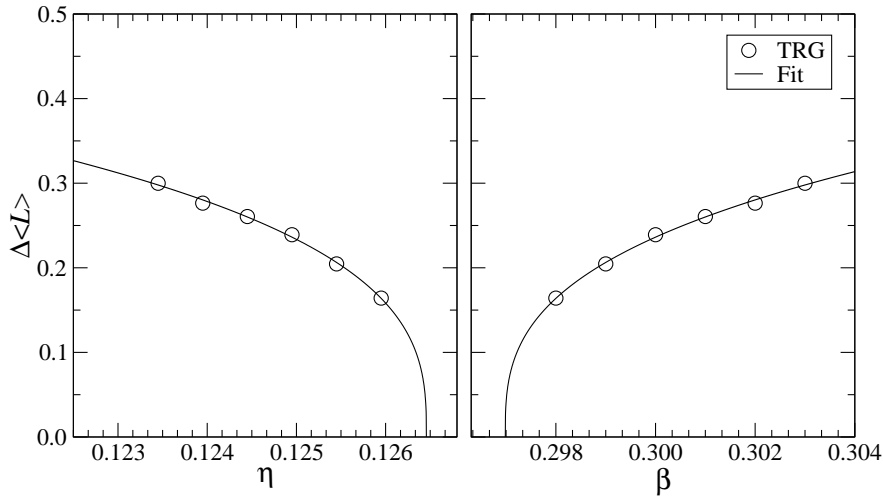}
	\caption{(Left) Fit of $\Delta \langle L\rangle$ at $\mu=2$ as a function of $\eta$. (Right) Fit of $\Delta \langle L\rangle$ at $\mu=2$ as a function of $\beta$.}
  	\label{fig:delta_4d_mu2_fit}
\end{figure}

\begin{figure}[htbp]
	\centering
	\includegraphics[width=1.0\hsize]{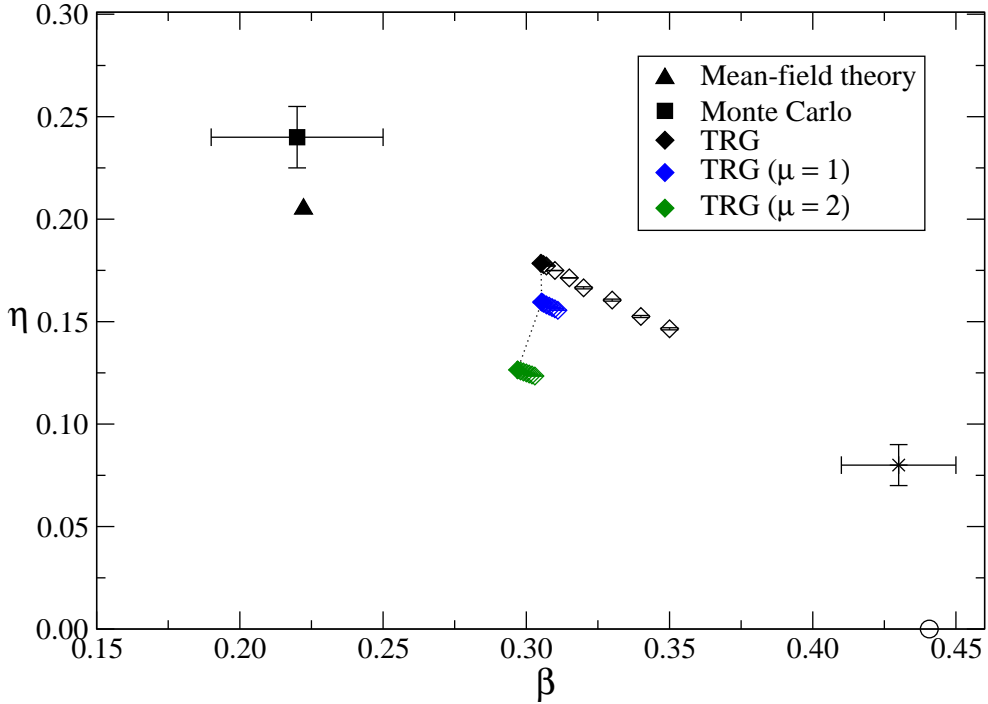}
	\caption{Summary of the critical points obtained by several methods on the (3+1)$d$ $\mathds{Z}_{2}$ gauge-Higgs model. The solid symbols correspond to the critical endpoints obtained by the mean-field theory~\cite{Brezin:1981zs} (triangle), the Monte Carlo simulation~\cite{Creutz:1979he} (square), and the TRG method with $D=48$ (diamond). All the dark symbols are located on the usual $\beta$-$\eta$ plane with vanishing $\mu$. Open diamonds show the first-order transition points located by the TRG method at each $\mu$. Blue diamonds show the phase transitions between the confinement and the Higgs phases at $\mu=1$ and green ones correspond to those at $\mu=2$. For reference, the triple point obtained by the Monte Carlo simulation~\cite{Creutz:1979he} is represented by the star symbol. Also, the open circle on the $\beta$-axis is the first-order pure gauge transition derived in Ref.~\cite{Balian:1974ir}.}
  	\label{fig:phasedgm_mu}
\end{figure}

Let us turn to the finite density cases with $\mu=1$ and $\mu=2$.  In Figs.~\ref{fig:link_mu=1} and \ref{fig:link_mu=2}, we plot the $\eta$ dependence of the link average with the several choices of $\beta$. Table~\ref{tab:link_4d} summarizes the finite values of $\Delta \langle L\rangle$ and the transition points. $\Delta \langle L\rangle$ is fitted with the same functions as in the case of $\mu=0$. The fit results are shown in Fig.~\ref{fig:delta_4d_mu1_fit} for $\mu=1$ and in Fig.~\ref{fig:delta_4d_mu2_fit} for $\mu=2$. Their numerical values are presented in Table~\ref{tab:link_4d_fit}, together with the result of $\mu=0$.
We obtain $(\beta_{\rm c},\eta_{\rm c})=(0.3053(2),0.1595(3))$ and (0.2969(2),0.1264(1)) as the critical endpoints at $\mu=1$ and $\mu=2$, respectively. Comparing the critical endpoints at $\mu=0$, $1$, $2$, we find that $\beta_{\rm c}$ has little $\mu$ dependence, while $\eta_{\rm c}$ is sizably diminished as $\mu$ increases. 
We summarize these findings in Fig.~\ref{fig:phasedgm_mu}, where we plot the critical endpoints determined by the TRG method comparing them with those obtained by other approaches~\cite{Brezin:1981zs,Creutz:1979he} and some other transition points such as the triple point~\cite{Creutz:1979he} and the pure gauge transition point~\cite{Balian:1974ir}. At $\mu=0$, it seems that the Monte Carlo calculation~\cite{Creutz:1979he} and the TRG one in this work shares a similar first-order line, though their resulting endpoints are different.


\begin{figure}[htbp]
	\centering
	\includegraphics[width=0.8\hsize]{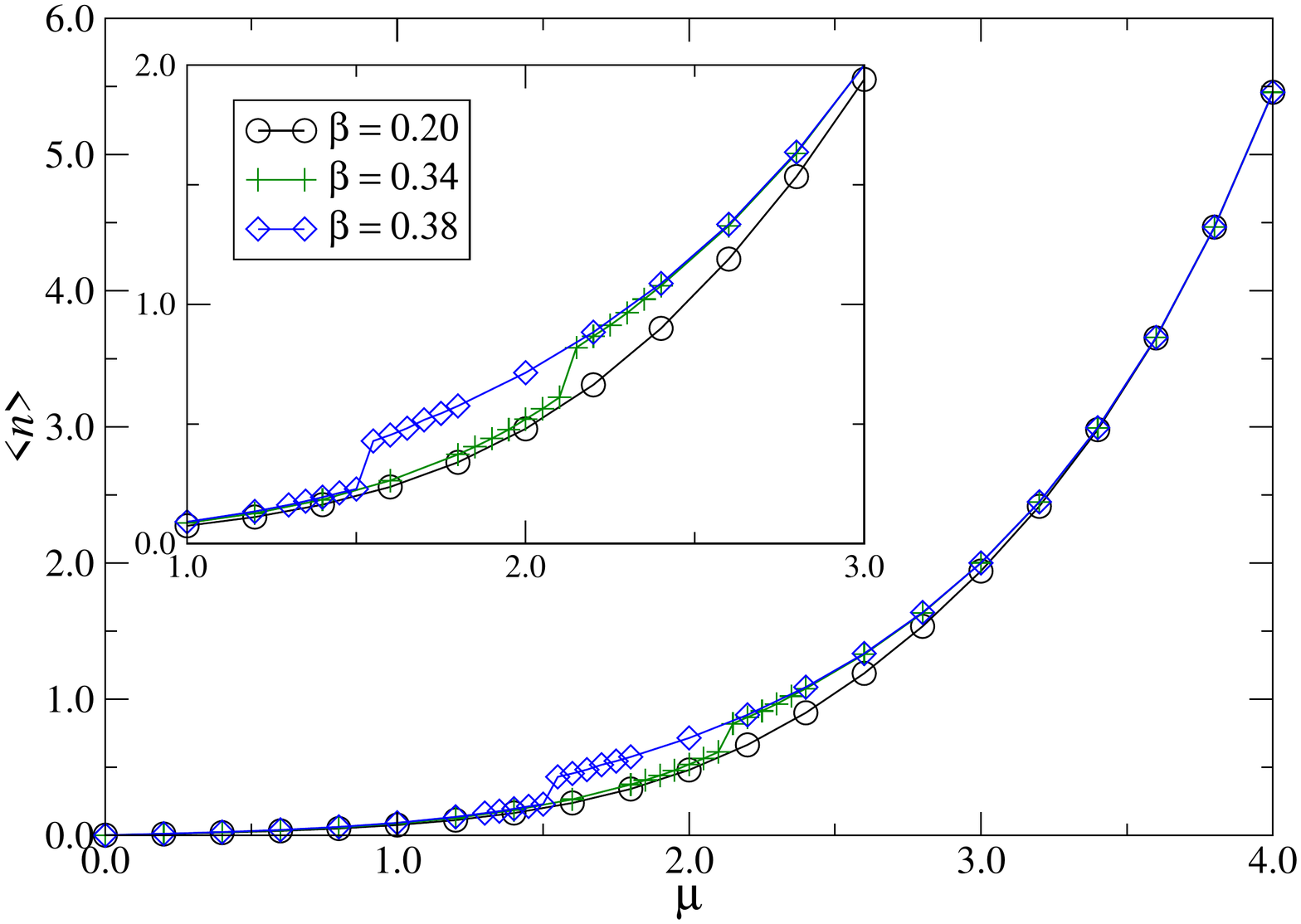}
	\caption{$\mu$ dependence of $\langle n\rangle$ at $\eta=0.1$ with $\beta=0.20$, $0.34$ and $0.38$. Inset graph magnifies the vicinity of the transition points at $\beta=0.34$ and $0.38$.}
  	\label{fig:n_mu}
\end{figure}

Finally, we investigate the $\mu$ dependence of the number density defined by
\begin{align}
\label{eq:number}
	\langle n\rangle=\frac{1}{V}\frac{\partial\ln Z}{\partial \mu},
\end{align}
which is also evaluated by the impurity tensor method.
In Fig.~\ref{fig:n_mu}, we plot the number density $\langle n\rangle$ as a function of $\mu$ with three choices of $\beta$ at $\eta=0.1$. We expect the confinement phase over $0\le \mu\le 4$ at $\beta=0.20$. 
At $\beta=0.34$ and $0.38$, the number density shows a finite gap at a certain point of $\mu$, which indicates that there exists the first-order phase transition from the confinement phase to the Higgs phase.

\section{Summary and outlook} 
\label{sec:summary}

This work is the first application of the TRG method to a four-dimensional lattice gauge and serves as a preparatory study for future investigation of the critical endpoint of the finite density QCD. 

We have investigated the critical endpoints of the higher-dimensional (more than two-dimensional) $\mathds{Z}_2$ gauge-Higgs model at finite density. 
To locate them, we have employed the average link $\langle L\rangle$ as an indicator: the critical endpoint is determined by vanishing $\Delta\langle L\rangle$.
In the (2+1)$d$ model, it has been confirmed that the resulting location of the critical endpoint at vanishing density is consistent with the recent result provided in Ref~\cite{Somoza:2020jkq}. Also, we find that the first-order transition points located by the TRG method are in excellent agreement with the self-dual line.
In the (3+1)$d$ model, the critical endpoints at $\mu=0$, $1$, $2$ are determined by the TRG calculation with $D=48$. Current results show that the critical inverse gauge coupling $\beta_{\rm c}$ has little $\mu$ dependence, while the critical spin-spin coupling $\eta_{\rm c}$ is sizably diminished as $\mu$ increases. 
At vanishing density, our estimation of the critical endpoint is inconsistent with the known estimated by the mean-field theory and the Monte Carlo studies.

The current study shows that the TRG method enables us to locate the critical endpoint investigating a certain observable along the first-order transition line. As a possible future work, it must be interesting to locate the triple point for this model and investigate the universality class as discussed in Refs.~\cite{Somoza:2020jkq,bonati2021multicritical}.  Although we have just focused on the simplest gauge group $\mathds{Z}_{2}$ and the model does not suffer from the sign problem, our strategy is easily extended to the $\mathds{Z}_{N}$ gauge-Higgs model with $N>2$ in arbitrary dimension. 
Since the TRG does allow us to study the systems with the sign problem even in four dimensions, as demonstrated by some practical computations in Refs.~\cite{Akiyama:2020ntf,Akiyama:2020soe}, we expect that the TRG is a promising method to investigate the higher-dimensional lattice gauge theories with the sign problem. This is a possible research direction as future work. 
As a next step, in addition, this study should be extended to the higher-dimensional lattice gauge theories with continuous gauge groups, also including dynamical matter fields.

\begin{acknowledgments}
  Numerical calculation for the present work was carried out with the supercomputer Fugaku and Oakforest-PACS (OFP) provided by RIKEN and JCAHPC, respectively, through the HPCI System Research Project (Project ID: hp210074, hp210204). We also used computational resources of OFP and Cygnus under the Interdisciplinary Computational Science Program of Center for Computational Sciences, University of Tsukuba.
This work is supported in part by Grants-in-Aid for Scientific Research from the Ministry of Education, Culture, Sports, Science and Technology (MEXT) (No. 20H00148) and JSPS KAKENHI Grant Number JP21J11226 (S.A.).
\end{acknowledgments}


\appendix

\section{Impurity tensor method}
\label{sec:impure}

We describe $\langle L\rangle$ defined in Eq.~\eqref{eq:def_al} by a sum of tensor networks each of which includes a single impurity. Eq.~\eqref{eq:def_al} is equivalent to
\begin{align}
\label{eq:imp_avl}
	(d+1)\langle L\rangle=\sum_{i=1}^{d}\frac{\langle U_{i}(n)\rangle}{V}+\cosh\mu\frac{\langle U_{d+1}(n)\rangle}{V}.
\end{align}
Assuming a link $\ell'=(p,\lambda)$ is included in a basic cell $c'$, we have
\begin{align}
\label{eq:u_lp}
	\langle U_{\lambda}(p)\rangle=\sum_{c'}\left(\frac{1}{Z}{\rm tTr}\left[\tilde{\mathcal{T}}_{c'}\prod_{c\neq c'}\mathcal{T}_{c}\right]\right),
\end{align}
where the impurity tensor $\tilde{\mathcal{T}}_{c'}$ is constructed in the almost same way with $\mathcal{T}_{c}$, just replacing the ``pure'' link tensor $A_{\ell'}$ by the following ``impure'' link tensor,
\begin{align}
\label{eq:impure_link}
	\tilde{A}_{\ell';m_{1}\cdots m_{d}m'_{1}\cdots m'_{d}}=\sum_{U_{\lambda}(p)}U_{\lambda}(p)\left(\prod_{j=1}^{d}V_{U_{\lambda}(p)m_{j}}V_{U_{\lambda}(p)m'_{j}}\right)M_{U_{\lambda}(p)}.
\end{align}
Thanks to a uniform structure of tensor network in Eq.~\eqref{eq:tn_lgt}, we can simplify Eq.~\eqref{eq:u_lp} by
\begin{align}
	\frac{\langle U_{\lambda}(p)\rangle}{V}=\frac{1}{Z}{\rm tTr}\left[\tilde{\mathcal{T}}_{c'}\prod_{c\neq c'}\mathcal{T}_{c}\right],
\end{align}
and Eq.~\eqref{eq:imp_avl} is finally expressed as
\begin{align}
	(d+1)\langle L\rangle=\frac{d}{Z}{\rm tTr}\left[\tilde{\mathcal{T}}_{s'}\prod_{c\neq s'}\mathcal{T}_{c}\right]+\frac{\cosh\mu}{Z}{\rm tTr}\left[\tilde{\mathcal{T}}_{t'}\prod_{c\neq t'}\mathcal{T}_{c}\right],
\end{align}
with two kinds of impurity tensors: $\tilde{\mathcal{T}}_{s'}$ includes an impure spatial link tensor and $\tilde{\mathcal{T}}_{t'}$ does an impure temporal link tensor.

Similarly, we can easily describe $\langle n\rangle$ defined in Eq.~\eqref{eq:number} by a tensor network just including a temporal impurity such that
\begin{align}
	\langle n\rangle=\frac{2\eta\sinh\mu}{Z}{\rm tTr}\left[\tilde{\mathcal{T}}_{t'}\prod_{c\neq t'}\mathcal{T}_{c}\right].
\end{align}


\bibliographystyle{JHEP}
\bibliography{bib/formulation,bib/algorithm,bib/discrete,bib/grassmann,bib/continuous,bib/gauge,bib/review,bib/for_this_paper}

\end{document}